# Specific Heat and Upper Critical Field of $Sc_5Ir_4Si_{10}$ Superconductor

Yue Sun[1,2], Yi Ding[1], Dongmei Gu[1], Jinchen Zhuang[1], Zhixiang Shi[1*], Tsuyoshi Tamegai[2]

[1]*Department of Physics, Southeast University, Nanjing 211189, People's Republic of China*

[2]*Department of Applied Physics, The University of Tokyo, 7-3-1 Hongo, Bunkyo-ku, Tokyo 113-8656, Japan*

## *Abstract*

Temperature and magnetic field dependent specific heat of textured $Sc_5Ir_4Si_{10}$ superconductor was investigated in detail. Based on the fitting of zero-field electronic specific heat by different gap structures as well as the discussion on field-induced specific heat coefficient, $Sc_5Ir_4Si_{10}$ is proved to be an anisotropic *s*-wave superconductor with a gap anisotropy, $\Delta_{min}/\Delta_{max}$, of 0.53. The anisotropy of upper critical field suggests the weakly one-dimensional Fermi surface, and the value is consistent with the gap anisotropy result obtained from the specific heat data.

# 1. Introduction

Recently discovered iron-based superconductors[1] have attracted much attention for some of their interesting properties, of which the interplay between superconductivity (SC) and some other ordered phases (spin-density wave (SDW),[2] orbital ordering,[3] antiferromagnetism (AFM)[4] and ferromagnetism (FM)[5]) has become a central issue and believed to be related to the origin of the high temperature superconductivity. Before this discovery, ternary silicide $R_xT_yX_z$ (R: rare earth element; T: transition metal; X: Si or Ge) was also concerned for their coexistence of multiple ordered phases, such as charge-density wave (CDW), SDW, AFM and SC.[6, 7] There are various phases of ternary silicide $R_xT_yX_z$ with different structures. Among these compounds, $R_2Fe_3Si_5$ are superconducting or magnetic, and $Lu_2Fe_3Si_5$ has the highest $T_c$ (6.1 K)[8] of iron-based superconductors before the discovery of iron arsenide superconductors. Recently, Nakajima *et al.*[9] reported that $Lu_2Fe_3Si_5$ is a typical two-gap superconductor and no magnetic order was found. Another important ternary silicide is the $R_5T_4X_{10}$,[10] in which the T and X atoms form planer nets of pentagons and hexagons that stack perpendicular to the basal plane of the primitive tetragonal cell, and are connected along the c axis via T-X-T zigzag chains. The layers of R atoms separate the pentagon-hexagon layers, and so the structure is characterized by the absence of T-T bonds. Although compounds belonging to the $R_5Ir_4Si_{10}$ family exhibit a variety of ground states depending on the nature of the rare-earth element R,[6] they are characterized by the CDW formation except for $Sc_5Ir_4Si_{10}$.[11] $Sc_5Ir_4Si_{10}$ shows superconductivity with the highest transition temperature similar to the iron-based superconductors, where the superconductivity emerges and moves to high temperature accompanied by the suppression of the SDW.[2] Thus the clarification of the paring mechanism as well as the electronic behavior of $Sc_5Ir_4Si_{10}$ is of great importance to the understanding of the interplay between SC and ordered phases. Unfortunately, as far as we know little effort has been done on this issue. Tamegai *et al.*[12] suggested that $Sc_5Ir_4Si_{10}$ may be a two-gap superconductor or only with weak gap anisotropy, while the intrinsic pairing mechanism and gap structure are still unknown.

In this paper, we address these issues mentioned above by the temperature and magnetic field dependent specific heat measurements as well as the study of the upper critical field, which confirms that $Sc_5Ir_4Si_{10}$ is an anisotropic *s*-wave superconductor.

## 2. Experimental Methods

Buttons of $Sc_5Ir_4Si_{10}$ were prepared by arc melting 5:4:10 stoichiometric mixtures of Sc, Ir and Si in an Ar environment. By cutting the buttons along certain direction, textured sample was obtained. Crystal structure were characterized by x-ray diffraction (XRD).Details of the preparation, structure analysis, transport and magnetic properties of the textured sample have been reported previously.[13] Magnetization measurements were performed using a commercial SQUID magnetometer (MPMS-XL5, Quantum Design). Specific heat data were obtained on the same sample under the applied field from 0 to 2 T using the Quantum Design Physical Property Measurement System (PPMS) via the relaxation method.

## 3. Results and Discussion

Inset of Fig. 1 shows XRD pattern of the textured $Sc_5Ir_4Si_{10}$ superconductor. It can be observed that there are only a few characteristic peaks, of which the highest two peaks correspond to that of (2, 2, 0) and (4, 4, 0), which means the large surface of this sample is (1, 1, 0) plane and the *c*-axis is parallel to the large surface. Detailed structure and composition analyses have been reported in our previous publication.[13] Temperature dependences of zero-field-cooled (ZFC) and field-cooled (FC) magnetizations at 10 Oe are shown in the main panel of Fig. 1. The textured $Sc_5Ir_4Si_{10}$ displays a $T_c \sim 8$ K with a very sharp transition width around 0.2 K (obtained from the criteria of 10 and 90% of the magnetization result at 2 K). The sharp transition width manifests the high quality of the texture sample.

Fig. 2 shows the specific heat divided by temperature $C/T$ as a function of $T^2$ for the textured $Sc_5Ir_4Si_{10}$ superconductor. A clear jump associated with the superconducting transition is observed at about 8 K under zero field, which is the same as that obtained by magnetization measurement. The superconducting jump is suppressed gradually with increasing magnetic field parallel to *ab*-plane, and there is no obvious sign of superconductivity when applied field is above 1.0 T. All the normal state data at various magnetic fields merge into a single straight line, which means the sample is free of magnetic impurities. This ensures the reliability of the obtained electronic behavior as well as the pairing mechanism below. The normal state specific heat divided by temperature can be fitted by the sum of electronic part ($\gamma_n$) and phononic part ($\beta_n T^2$) with its high order term ($\alpha_n T^4$): $C_n/T = \gamma_n + \beta_n T^2 + \alpha_n T^4$. The fitting result was shown as the solid line in Fig. 2 giving $\gamma_n$ = 32.4 mJ/molK$^2$, $\beta_n$ = 0.72 mJ/molK$^4$, $\alpha_n$ = 5.91× 10$^{-4}$ mJ/molK$^6$, which are close to that reported in the single crystal.[12] By using the relation: $\Theta_D = (\frac{12\pi^4 N_A Z k_B}{5\beta_n})^{1/3}$, where $N_A$ is the Avogadro constant, Z is the number of atoms per formula unit, and $k_B$ is the Boltzmann's constant, we obtain the Debye temperature $\Theta_D$ = 371.5 K. With the value of $\Theta_D$ and $T_c$, the electron-phonon coupling parameter $\lambda_{ep}$ can be estimated from the McMillan theory, where $\lambda_{ep}$ is given by $\lambda_{ep} = \frac{1.04 + \mu^* \ln(\Theta_D/1.45T_C)}{(1-0.62\mu^*)\ln(\Theta_D/1.45T_C) - 1.04}$. With $\mu^*$ = 0.1, we find the value of $\lambda_{ep}$ to be 0.63, indicating a moderate coupling strength. Inset of Fig. 2 depicts the specific heat divided by temperature *C/T* as a function of *T*, from which we can easily obtain the specific jump at $T_c$. Together with the obtained Sommerfeld coefficient $\gamma_n$, the value of the normalized jump of the specific heat at $T_c$, $\Delta C/\gamma_n T_c$, is obtained as 1.75 larger than the BCS value of 1.43. Until now without any model fitting, the values of normalized specific heat jump and the electron-phonon coupling constant have already implied a stronger coupling strength in Sc$_5$Ir$_4$Si$_{10}$.

Zero field electronic specific heat $C_e/\gamma_n T_c$ was obtained from the subtraction of phonon terms, and fitted by the following formula[14] based on the BCS theory

$$C_e = 2N(0)\beta k \frac{1}{4\pi}\int_0^{2\pi} d\phi \int_0^{\pi} d\theta \sin\theta \int_{-\hbar\omega_D}^{\hbar\omega_D} -\frac{\partial f}{\partial E}(E^2 + \frac{1}{2}\beta\frac{d\Delta^2}{d\beta})d\varepsilon, \qquad (1)$$

where $N(0)$ is the density of states at the Fermi surface, $\beta = 1/k_B T$, $E = (\varepsilon^2 + \Delta^2)^{1/2}$, $\Delta = \Delta_0$

the superconducting energy gap for isotropic *s*-wave, $\Delta = \Delta_0 \cos n\phi$ for line-node (n=2 (*d*-wave) for calculation), $\Delta = \Delta_0 \sin n\theta$ for point-node (n=1 for calculation). The best fitting gives $2\Delta/k_B T_c = 5.63 \pm 0.05$ for line-node, $2\Delta/k_B T_c = 4.79 \pm 0.03$ for point-node, and $2\Delta/k_B T_c = 3.80 \pm 0.003$ for *s*-wave, which are plotted in Figs. 3 (a) – (c), respectively. Obviously the *s*-wave fitting is much better than the line-node and point-node models, although there are still some deviations from the experimental data especially at low temperatures. It seems that the gap structure of $Sc_5Ir_4Si_{10}$ more likely to be nodeless *s*-wave, but it cannot be simply explained by the isotropic *s*-wave. Thus the two-gap and anisotropic *s*-wave models are further applied to fit the experimental data.

The two-gap model, which has been proved correct for another intermetallic compound $Lu_2Fe_3Si_5$[9] and may be applicable also in $Sc_5Ir_4Si_{10}$,[12] contains two distinct gaps, $\Delta = \gamma_1\Delta_1 + \gamma_2\Delta_2$. The line shown in Fig. 3 (d) represents the fitting result with the magnitude of two distinct superconducting gaps, $2\Delta_1/k_B T_c = 4.01 \pm 0.08$ and $2\Delta_2/k_B T_c = 3.02 \pm 0.03$, with the relative weight of $\gamma_1 = 0.74 \pm 0.01$ and $\gamma_2 = 0.26 \pm 0.01$, respectively. The anisotropic *s*-wave model has been successfully used to explain the specific heat of some other superconductors, like CaAlSi, $LuNi_2B_2C$, and $YNi_2B_2C$[15), 16), 17]. For the anisotropic *s*-wave, the gap structure can be expressed by $\Delta = \Delta_0/\sqrt{1-\varepsilon\cos^2(\theta)}$, and the parameter $-\infty \leq \varepsilon \leq 1$ is related to eccentricity *e* as $\varepsilon = e^2 = 1 - c^{-1}$ where *c* is the normalized semiaxis length along the *c* axis.[15] The fitting result based on the anisotropic *s*-wave model with the magnitude of the gap $2\Delta/k_B T_c = 3.19 \pm 0.03$, and the parameter $\varepsilon = 0.72 \pm 0.04$, corresponding to the gap anisotropy $\Delta_{min}/\Delta_{max}$ of 0.53±0.04, is shown in Fig. 3 (e). In order to compare the quality of fitting results by different gap models, differences between the fitting results and the experimental data are shown in Fig. 3 (f), which clearly shows that the two-gap and anisotropic *s*-wave models can reproduce the experimental result better than the isotropic *s*-wave, especially at low temperatures. However, the differences of isotropic *s*-wave, two-gap and anisotropic *s*-wave models are relatively too small to distinguish the pairing mechanism of $Sc_5Ir_4Si_{10}$. Thus, more evidences are still needed to identify this problem.

Another approach to study the pairing mechanism is through the vortex excitation in the mixed state, which is also sensitive to the gap structure.[18] As for a conventional *s*-wave superconductor, the specific heat in the vortex state is dominated by the contribution from the localized quasiparticles in the vortex core, and the specific heat coefficient $\gamma(H)$ is in a linear relationship with the magnetic field. On the other hand, for a superconductor with nodes in the gap, for example *d*-wave superconductors with line nodes, Volovik *et al.*[19] pointed out that supercurrents around a vortex core in the mixed state cause a Doppler shift of the quasiparticles excitation spectrum. This shift has an important effect upon the low energy excitation around the nodes leading to a $H^{1/2}$ dependence of the coefficient $\gamma(H)$. We plotted the field dependence of field induced specific heat coefficient $\Delta\gamma(H) = [C(T,H) - C(T,0)]/T$ in Fig. 4. Obviously, $\Delta\gamma(H)$ versus field $H$ deviates from both the linear and square-root behavior. It proves $Sc_5Ir_4Si_{10}$ is not a simple isotropic *s*-wave or nodal superconductor in accordance with the fitting results of the zero-field specific heat.

In our case, the slope of $\Delta\gamma(H)$ at low field is larger than that at higher field, which has been witnessed in two-gap superconductors such as $MgB_2$[20] and $Lu_2Fe_3Si_5$[21], as well as anisotropic *s*-wave superconductors such as $LuNi_2B_2C$[16], $YNi_2B_2C$[17], and $Li_xZrNCl$[22]. For a typical two-gap superconductor, the $\Delta\gamma(H)$ first increases steeply with the field caused by the suppression of the smaller gap by increasing magnetic field. After the magnetic field increased above the virtual upper critical field, $H^*$, of the smaller gap, the specific heat coefficient will increase linearly with the applied field. Based on this, the virtual upper critical field $H^*$ is obtained as 0.17 T shown as the kink of the line in Fig. 4. In fact, $\Delta\gamma(H)$ at $H^*$ is about $0.34\gamma_n$. By subtracting the contribution from the large gap, obtained from the linear extrapolation of high-field $\Delta\gamma(H)$, $\Delta\gamma(H)$ for the smaller gap is calculated ~ $0.14\gamma_n$, which is different from $0.26\gamma_n$ obtained from the two-gap fitting of zero field specific heat. Furthermore, the virtual field $H^*$ and upper critical field provide us an simple way to estimate the ratio of the two gaps.[20, 21] The virtual upper critical field for the small gap can be written as $H^* \sim \Phi_0 / 2\pi \xi_{ab}^* \xi_c^*$, and the upper critical field for $H \parallel ab$ can be written as $H_{c2}^{ab} \sim \Phi_0 / 2\pi \xi_{ab} \xi_c$. Here, $\xi_{ab}^* = \hbar v_{F1}^{ab} / \pi \Delta_1$ and $\xi_c^* = \hbar v_{F1}^c / \pi \Delta_1$ are the coherence length along $ab$ plane and $c$-axis for the small gap. $\xi_{ab} = \hbar v_{F2}^{ab} / \pi \Delta_2$ and $\xi_c = \hbar v_{F2}^c / \pi \Delta_2$ are the coherence length along $ab$ plane and $c$-axis for the large gap. $v_{Fi}$ ($i$=1, 2) is the Fermi velocity for each band. Thus, the ratio of the two gaps can be estimated as $\Delta_2 / \Delta_1 = \sqrt{H_{c2}^* / H_{c2}^{ab}} \sqrt{(v_{F2}^{ab} v_{F2}^c) / (v_{F1}^{ab} v_{F1}^c)}$, where $H_{c2}^{ab}$ is ~ 11.7 kOe from the upper critical field fitting below. If we simply assume the Fermi velocities in the two bands are the same, the gap ratio can be calculated as 0.38, which is much smaller than the value ~ 0.75 from zero field specific heat fitting result. Based on the discussion above, $Sc_5Ir_4Si_{10}$ seems not a two-gap superconductor. In other words, $T_c$ of $Sc_5Ir_4Si_{10}$ is almost independent of nonmagnetic impurities doping,[23] quite different from the similar compound $Lu_2Fe_3Si_5$, which is already proved to be a two-gap superconductor,[9, 24] and $T_c$ is rapidly suppressed by nonmagnetic impurities.[25, 26] In the case of anisotropic $s$-wave, Nakai et al.[27] theoretically revealed that $\gamma(H)$ shows a crossover from $H$ linear dependence at low field to $\sqrt{H}$ at high fields. Thus we replot the $\Delta\gamma(H)$ as a function of $\sqrt{H}$ in the inset of Fig. 4. At low field $\Delta\gamma(H)$ increases with a upward curvature with respective to $\sqrt{H}$, while, at fields larger than 0.3 T, $\Delta\gamma(H)$ obviously increase linearly to $\sqrt{H}$, which is in accordance with Nakai's calculation result on anisotropic $s$-wave superconductors. Combined with the zero field specific heat fitting results and the field dependence of specific heat coefficient, $Sc_5Ir_4Si_{10}$ is an anisotropic $s$-wave superconductor.

Fig. 5 shows the temperature dependence of the upper critical field for $H // ab$ and $H // c$ from $M$-$T$ of this textured sample[13] together with that from $Sc_5Ir_4Si_{10}$ single crystal.[11] In both directions, $H_{c2}$ curves exhibit convex shapes flatten at low temperatures, which is strikingly different from the linear or sublinear increase with a concave shape in the two-band model.[21, 28, 29] Therefore, to describe the $H_{c2}$ curves in the present case, the commonly used Werthamer-Helfand-Hohenberg (WHH) theory[30] was adopted here. According to the WHH theory, $H_{c2}$ limited by the orbital depairing in the dirty limit is given by

$$H_{c2}^{orb}(0) = -0.69 T_c dH_{c2}/dT \big|_{T=T_c} \quad (2)$$

In this situation, the superconductivity is destroyed when the kinetic energy of the Cooper pair exceeds condensation energy. On the other hand, superconductivity is also suppressed while the magnetic energy associated with the Pauli spin susceptibility exceeds the condensation energy in the superconducting state as a result of Zeeman effect. Thus, in the fitting of $H_{c2}$ by WHH theory, we incorporate the spin paramagnetic effect via the Maki parameter, $\alpha$, and also the spin-orbit effect by the scattering constant $\lambda$.

For $H // c$, $H_{c2}$ obtained from both the textured sample and single crystal sample can be fitting by $\alpha = 0$, $\lambda = 0$ with the initial slope of 4.0 kOe/K and 3.2 kOe/K, respectively. The obtained value of $H_{c2}$ at 0 K is about 27.8 kOe and 22.1 kOe for the textured and single crystal sample, respectively. The textured sample has higher $H_{c2}(0)$ possibly due to the microstructure defects. On the other hand, for $H // ab$, experimental data is below the calculated curves with $\alpha = 0$, $\lambda = 0$, which means the spin paramagnetic effect cannot be ignored. By introducing $\alpha = 2.4$, $\lambda = 10$, $H_{c2}$ obtained from both the textured sample and single crystalline samples can be well fitted with the initial slope of 2.0 kOe/K and 1.6 kOe/K, giving the $H_{c2}(0)$ values of 11.7 kOe and 9.5 kOe, respectively.

In the conventional BCS superconductors, the Pauli-limited upper critical field, $H_p$ can be simply estimated by $\mu_0 H_p$ [Tesla] $= 1.84 T_c$ [K], which is about 147 kOe in $Sc_5Ir_4Si_{10}$ superconductor. The value of $H_p$ is much larger than the orbital depairing limited upper critical field, $H_{c2}^{orb}(0)$, obtained above, which implies the spin paramagnetic effect may not play an important role in $H_{c2}$ of $Sc_5Ir_4Si_{10}$. Furthermore, with the value of $H_p$ and $H_{c2}^{orb}$, Maki parameter $\alpha$ can be estimated by $\alpha = \sqrt{2} H_{c2}^{orb} / H^p$. The calculated value of $\alpha$ is just 0.27, which is much smaller than that of 2.4 obtained from the fitting by WHH theory. Thus, the suppression of $H_{c2}$ for $H // ab$ may just come from the gap anisotropy rather than the spin paramagnetic effect.

The anisotropy of upper critical field $\gamma = H_{c2}^{ab} / H_{c2}^{c}$ as a function of temperature is shown in the inset of Fig. 5. $\gamma$ is almost temperature independent except for the 6 and 7 K data of the textured sample, which may come from the slight difference in sample quality.[13] The almost constant value of $\gamma$ is quite different from the strongly temperature dependent anisotropy in the multiband superconductors due to different bands' anisotropies, such as $MgB_2$ and iron-based superconductors.[31, 32] This again supports that $Sc_5Ir_4Si_{10}$ is not a two-gap superconductor. The value of $\gamma$ is roughly 0.45, reflecting the weakly one-dimensional Fermi surface, which is in consistent with the chainlike structure along the *c*-axis and the quasi-one-dimensional electronic band structure.[11, 12, 33, 34] Actually, if we assume the Fermi velocities parallel to *c*-axis and *ab* plane are similar, according to the Ginzburg-Landau (GL) theory,[35, 36] $\gamma$ at $T_c$ is determined by the gap anisotropy. Thus, the gap anisotropy can be estimated as 0.45, which is in consistent with the value of 0.53 obtained from the specific heat.

## 4. Conclusion

In summary, we have investigated the specific heat and upper critical field of $Sc_5Ir_4Si_{10}$ superconductor. The fitting results of the zero field electronic specific heat on different gap structures, and the discussion of field-induced electronic specific coefficient confirmed that $Sc_5Ir_4Si_{10}$ is an anisotropic *s*-wave superconductor with the gap anisotropy $\Delta_{min}/\Delta_{max}$ of 0.53. The upper critical field was studied in detail by the WHH model including the spin paramagnetic and spin-orbit effect. The anisotropy $\gamma$ of the upper critical field is about 0.45, which suggests the weakly one-dimensional Fermi surface. The value of $\gamma$ is also in consistent with the gap anisotropy obtained from the specific heat data.

## Acknowledgments

This work was supported by the Natural Science Foundation of China, the Ministry of Science and Technology of China (973 project: No. 2011CBA00105), Scientific Research Foundation of Graduate School (Grant No. YBJJ1104) of Southeast University, and Scientific Innovation Research Foundation of College Graduate in Jiangsu Province (CXZZ_0135).

*<u>zxshi@seu.edu.cn</u>

**Figure captions**

Fig. 1: (Color online) Temperature dependences of zero-field-cooled (ZFC) and field-cooled (FC) magnetizations at 10 Oe for $Sc_5Ir_4Si_{10}$. The inset shows XRD pattern of the textured sample.

Fig. 2: (Color online) Temperature dependence of specific heat plotted as $C/T$ vs $T^2$ at various magnetic fields for $Sc_5Ir_4Si_{10}$. The solid line represents the fit to the data in the normal state $C_n/T = \gamma_n + \beta_n T^2 + \alpha_n T^4$. The inset shows the specific heat jump at zero field plotted as $C/T$ vs $T$.

Fig. 3: (Color online) Various fitting of normalized electronic specific heat $C_e/\gamma_n T_c$ vs $T$ using (a) line nodes, (b) point nodes, (c) isotropic $s$-wave, (d) two-gap, (e) anisotropic $s$-wave models. (f) Differences between the fitting results and the experiment data (DF) for different models.

Fig. 4: (Color online) Magnetic field dependence of field-induced specific heat coefficient $\Delta\gamma(H)$. The inset plots the relationship of $\Delta\gamma(H)$ vs $H^{0.5}$.

Fig. 5: (Color online) Temperature dependence of upper critical fields of the textured and single crystalline $Sc_5Ir_4Si_{10}$. Solid lines are fits to the WHH theory without considering the spin paramagnetic effect. Dashed lines are the fitting curves by WHH model including the spin paramagnetic ($\alpha = 2.4$) and spin-orbit ($\lambda = 10$) effect. The inset is the anisotropy of upper critical field in $Sc_5Ir_4Si_{10}$.

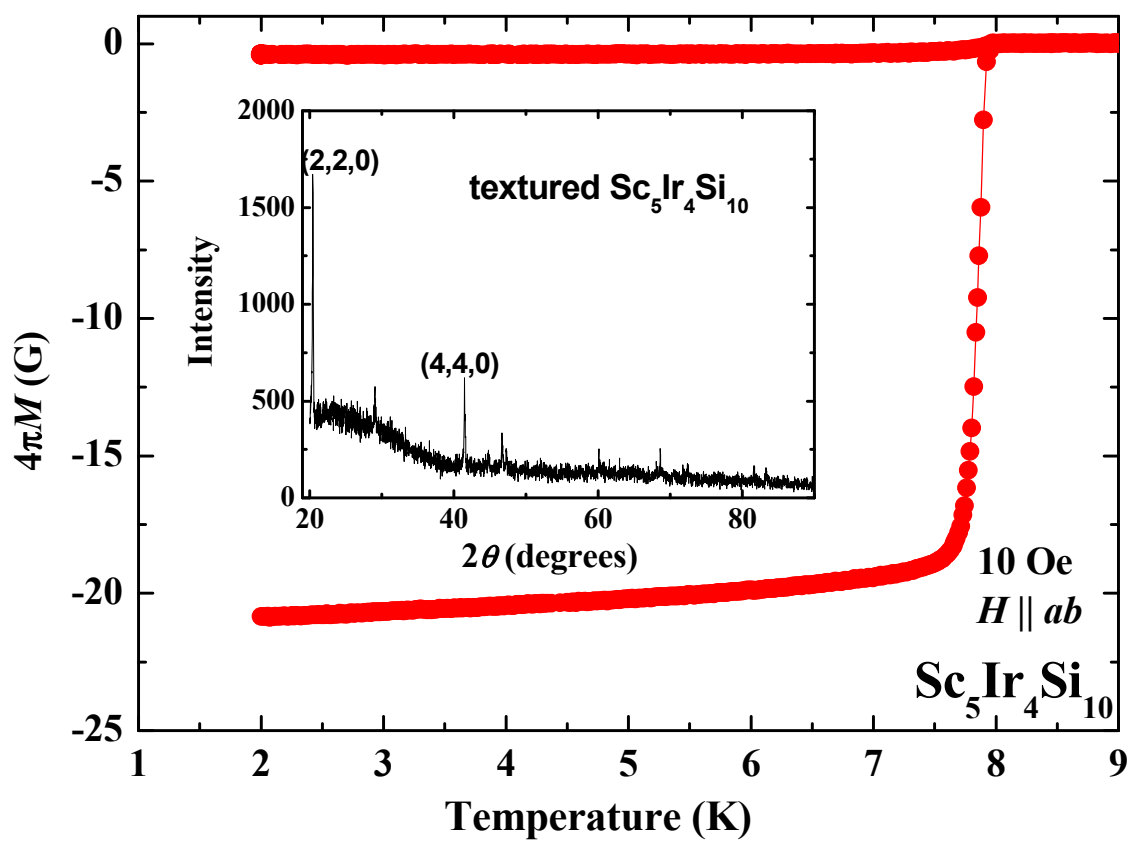

**Fig. 1** Yue Sun *et al.*

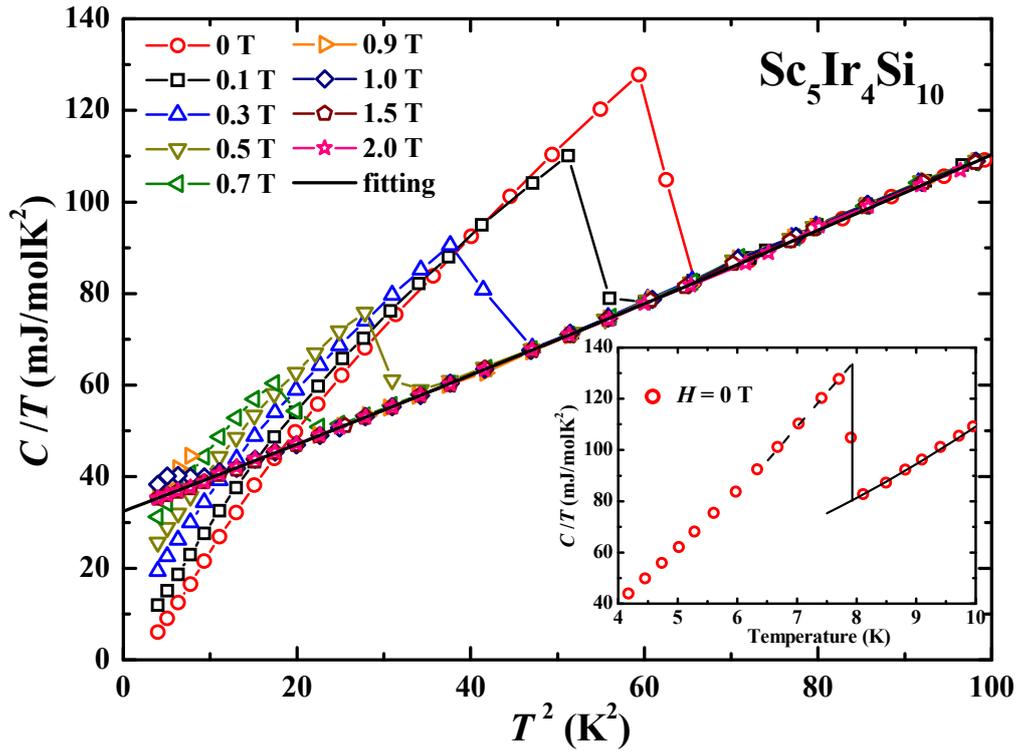

**Fig. 2 Yue Sun *et al.*** 

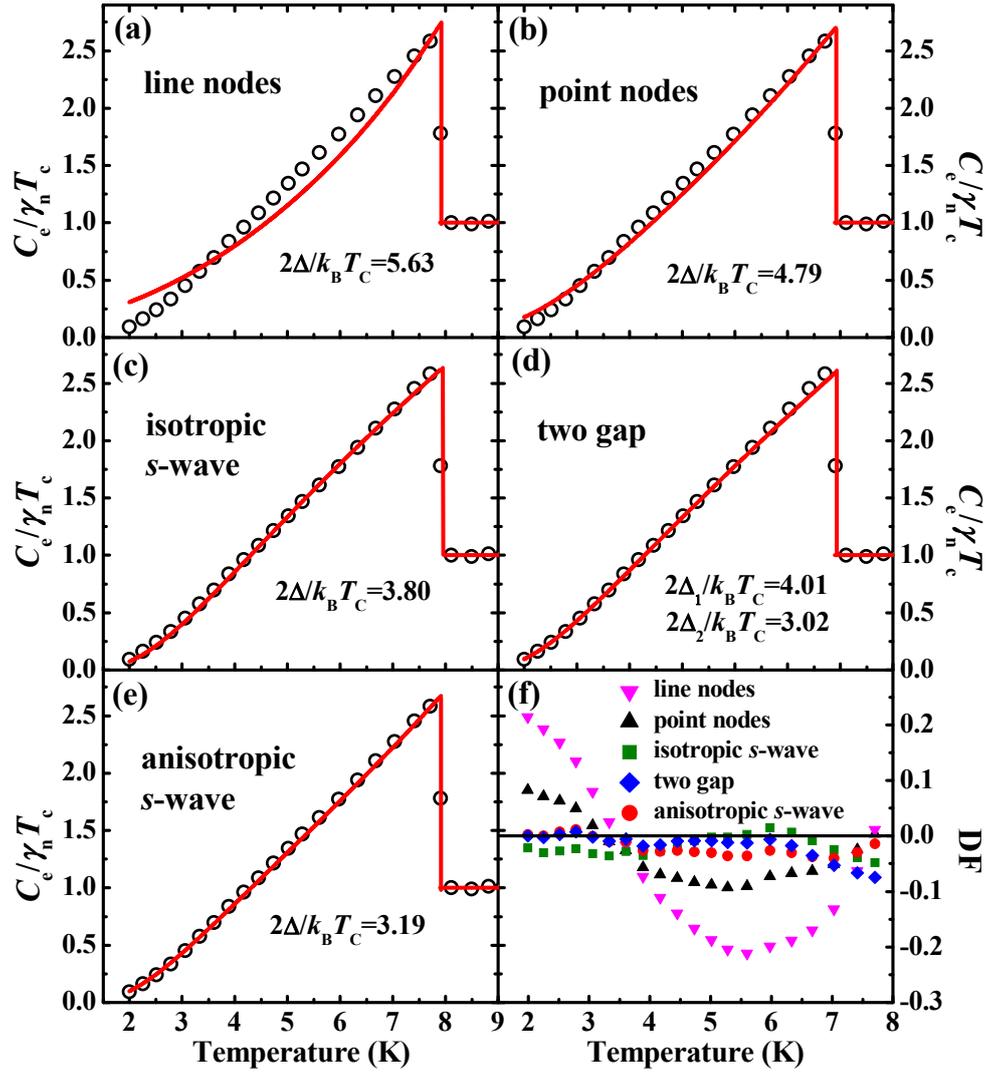

**Fig. 3** Yue Sun *et al.*

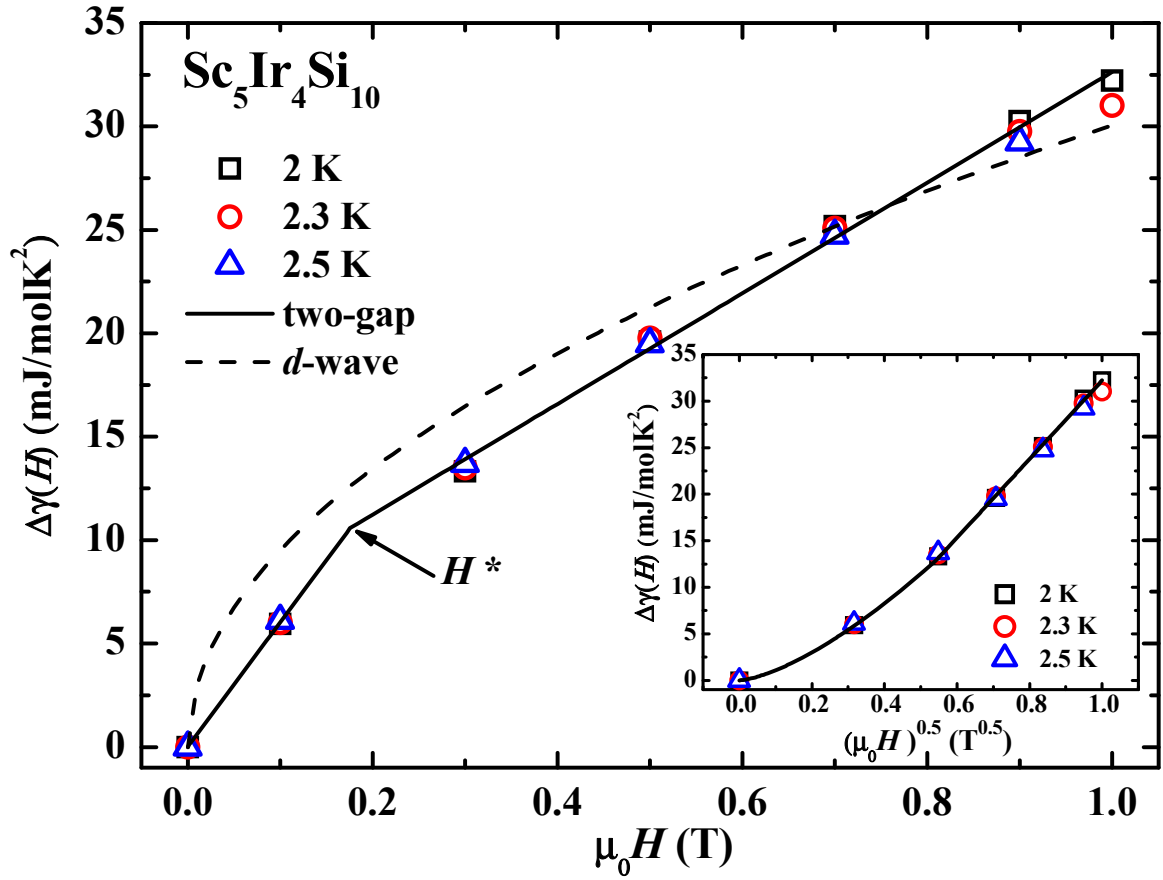

Fig. 4 Yue Sun *et al.*

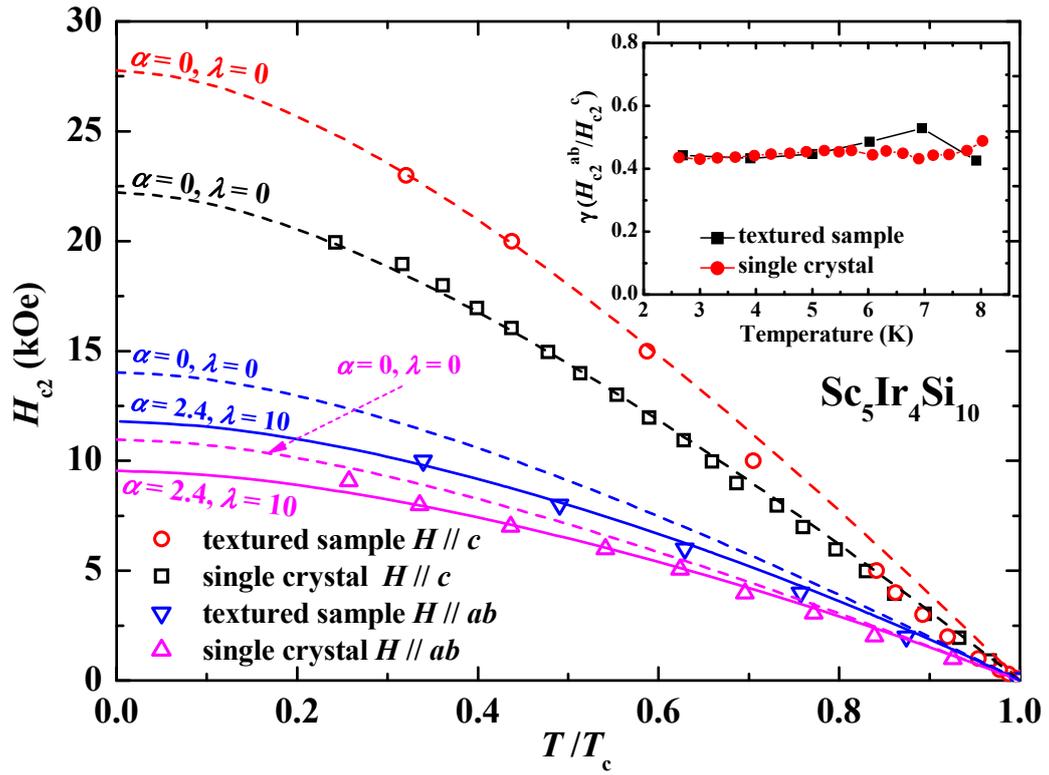

**Fig. 5 Yue Sun *et al.***